\documentclass[aps,prd,preprint,superscriptaddress,amsmath,amssymb,showpacs]{revtex4-1}
\usepackage{dcolumn}
\usepackage{graphicx}
\usepackage{float}
\usepackage{physics}
\usepackage[colorlinks=true,allcolors=blue]{hyperref}
\usepackage{mathrsfs}

\begin{document}

\title{A systematically study of thermal width of heavy quarkonia in a finite temperature magnetized background from holography}

\author{Sheng-Qin Feng}
\email{fengsq@ctgu.edu.cn}
\affiliation{College of Science, China Three Gorges University, Yichang 443002, China}
\affiliation{Key Laboratory of Quark and Lepton Physics (MOE) and Institute of Particle Physics,\\
Central China Normal University, Wuhan 430079, China}

\author{Yan-Qing Zhao}
\affiliation{College of Science, China Three Gorges University, Yichang 443002, China}

\author{Xun Chen}
\affiliation{Key Laboratory of Quark and Lepton Physics (MOE) and Institute of Particle Physics,\\
Central China Normal University, Wuhan 430079, China}
\date{\today}

\begin{abstract}
By simulating the finite temperatures magnetized background in the RHIC and LHC energies, we systematically study the characteristics of thermal widths and potentials of heavy quarkonia. It is found that the magnetic field has less influence on the real potential, but has a significant influence on the imaginary potential, especially in the low deconfined temperature. Extracted from the effect of thermal worldsheet fluctuations about the classical configuration, the
thermal width of $\Upsilon(1s)$ in the finite temperature magnetized background is investigated. It is found that at the low deconfined temperature the magnetic field can generate a significant thermal fluctuation of the thermal width of $\Upsilon(1s)$, but with the increase of temperature, the effect of magnetic field on the thermal width becomes less important, which means the effect of high temperature completely exceeds that of magnetic field and magnetic field become less important at high temperature.  The thermal width decreases with the increasing rapidity at the finite temperature magnetized background.  It is also observed that the effect of the magnetic field on the thermal width when dipole moving parallel to the magnetic field direction are larger than that moving perpendicular to the magnetic field direction, which implies that the magnetic field tends to enhance thermal fluctuation when dipole moving parallel to the direction of magnetic field. The thermal width of $\Upsilon(1S)$ hardly changes with the increasing temperature when dipole moving perpendicular to the magnetic field. But when dipole moving parallel to the magnetic field, the thermal width at low temperature is obviously larger than that at high temperature.
\end{abstract}

\keywords{thermal width, imaginary potential, finite temperature, magnetic field}

\pacs{11.25.Tq, 25.75.Nq}

\maketitle

\section{Introduction}\label{sec:01_intro}
 A new state of matter, so-called Quark-Gluon Plasma(QGP), has been generated in relativistic heavy ion collisions at the Relativistic Heavy Ion Collider (RHIC) and the Large Hadron Collider (LHC)~\cite{Adams:2005dq,Adcox:2004mh,Shuryak:2004cy}. The heavy quarkonia ($J/\Psi$ and $\Upsilon$ mainly) are formed in hard processes before the thermalization of the plasma, which are taken as useful probes to study the formation and evolution of the QGP\cite{Shuryak:1980tp}. The well-known work of Matsui and Satz \cite{Matsui:1986dk} argued that the binding interaction of the heavy quark-antiquark (Q$\bar{Q}$) pair in a thermal bath is screened by the medium, resulting in the melting of the heavy quarkonia. In the study of heavy ion collisions, besides high temperature, another important finding is the generation of a strong magnetic field of noncentral heavy ion collisions at the RHIC and the LHC\cite{Skokov:2009qp,Voronyuk:2011jd,Bzdak:2011yy,Deng:2012pc,Mo:2013qya,Zhong:2014cda,Zhong:2014sua,She:2017icp}. Since the magnetic field in relativistic heavy-ion collisions is so great that people believe that the strong magnetic fields can provide some deep investigations of the dynamics of quantum chromodynamics (QCD).

 The interaction energy $V_{Q\bar{Q}}$ of Q$\bar{Q}$ pair may possess a finite imaginary part $\textrm{Im}⁡V_{Q\bar{Q}}$, which can be used to calculate a thermal width of the quarkonium at finite temperature\cite{Gubser:1998bc,Aharony:1999ti,Laine:2007gj,Petreczky:2010tk}. The calculations of $\textrm{Im}⁡V_{Q\bar{Q}}$ related to heavy ion collisions in QCD have been carried out for static $Q\bar{Q}$ pairs by using lattice QCD\cite{Brambilla:2010vq,Burnier:2009yu,Rothkopf:2011db} and perturbative QCD\cite{Gubser:1998bc}. The dissociation of quarkonia is one of the most important experimental signal for QGP formation. Some publications argued\cite{Laine:2006ns,Beraudo:2007ky,Brambilla:2008cx,Escobedo:2014gwa} that the imaginary part of the potential $\textrm{Im}⁡V_{Q\bar{Q}}$ may be an important reason responsible for this suppression rather than color screening. The imaginary potential has been subsequently studied in weakly coupled theories by Refs.\cite{Brambilla:2010vq,Dumitru:2009fy}. However, an available method of the imaginary potential in recent years \cite{Gubser:1998bc,Margotta:2011ta,Maldacena:1997re} has be used in strongly coupled theories with the aid of nonperturbative methods of AdS/QCD.

 The imaginary potential of quarkonia for $\mathcal{N}$ = 4 SYM theory was studied by Noronha and Dumitru in their seminal work~\cite{Noronha:2009da}. This imaginary contribution originates from thermal fluctuations around the bottom of the classical sagging string in the bulk that links the heavy quarks situated at the boundary in the dual gravity picture. The imaginary potential $\textrm{Im}⁡V_{Q\bar{Q}}$  related to the effect of thermal fluctuations is due to the interactions between the heavy quarkonia and the medium.
Subsequently, a large number of research work about $\textrm{Im}⁡V_{Q\bar{Q}}$ were carried out with gauge/gravity duality. For instance, the $\textrm{Im}⁡V_{Q\bar{Q}}$ of static quarkonia was studied in \cite{Finazzo:2013aoa,Fadafan:2013bva}. Refs. \cite{Finazzo:2014rca,Ali-Akbari:2014vpa} studied the effect of moving quarkonia on $\textrm{Im}⁡V_{Q\bar{Q}}$. The influences of chemical potential and magnetic field on $\textrm{Im}⁡V_{Q\bar{Q}}$ were investigated in \cite{Zhang:2016tem,Zhang:2018fpe}. The studies of $\textrm{Im}⁡V_{Q\bar{Q}}$ in some AdS/QCD models are provided in \cite{Braga:2016oem,Sadeghi:2014zya}.

In the other hands, strong magnetic field  plays an essential roles in non-central heavy ion collisions at RHIC and LHC~\cite{Skokov:2009qp,Voronyuk:2011jd,Bzdak:2011yy,Deng:2012pc,Mo:2013qya,Zhong:2014cda,Zhong:2014sua,She:2017icp}. Strong magnetic field also provides a good probe of the dynamics of QCD. To accurately determine the suppression of quarkonia formed in relativistic heavy ion collisions, it is necessary to evaluate thermal width of $\Upsilon(1S)$ for moving quarkonia in the finite temperature magnetized QGP background in the RHIC and LHC energies.

Ref. \cite{Faulkner:2008qk} computed the momentum dependence of meson widths within the gauge/gravity duality. It was proposed that the thermal width becomes very large for rapidly moving meson, and the imaginary part of rapidly moving mesons may be already large enough to cause suppression of these states in a strongly coupled plasma even before complete dissociation. Thus, Refs. \cite{Liu:2006nn,Liu:2006he,Ejaz:2007hg} indicated that the dissociation temperature of meson decreased with the pair’s rapidity.
By simulating the background of finite temperatures magnetized background in the RHIC and LHC energies, We restrict ourselves to the range of temperature and magnetic field corresponding to RHIC and LHC energy regions to study the potential and thermal width for dipole moving parallel and pedicular to the magnetic field.
This paper is organized as follows: in Sec.~\ref{sec:02}, we introduce the setup of the gravity background with back-reaction of magnetic field through the Einstein- Maxwell (EM) system. These cases where the $Q\bar{Q}$ dipole moving parallel and perpendicular to the direction of the magnetic field are discussed in Sec.~\ref{sec:03} and Sec.~\ref{sec:04}, respectively. In Sec.~\ref{sec:05} we make a comparison of the results of dipole moving parallel and perpendicular to the magnetic field direction. And then we make conclusions in Sec.~\ref{sec:06}.

\section{The setup}\label{sec:02}
The action of the gravity background with back-reaction of magnetic field through the Einstein-Maxwell (EM) system\cite{Mamo:2015dea,Dudal:2015wfn,Li:2016gfn} is given as:
\begin{equation}
\label{eq1}
\ S = \frac{1}{16\pi G_5 }\int \dd{x}^5{ \sqrt{- g }(R-F^{MN}F_{MN}+\frac{12}{L^{2} })},
\end{equation}

\noindent where $R$ is the scalar curvature, $G_{5}$ is $5D$ Newton constant, $g$ is the determinant of metric $g_{\mu\nu}$, $L$ is the AdS radius and $F_{MN}$ is the U(1) gauge field \cite{Li:2016gfn}.

The Einstein equation for the EM system could be derived as follows
\begin{gather}
\label{eq2}
\ E_{MN}-\frac{6}{L^2}g_{MN}-2(g^{IJ}F_{MJ}F_{NJ}-\frac{1}{4}F_{IJ}F^{IJ}g_{MN})=0,
\end{gather}

\noindent where $E_{MN}=R_{MN}-\frac{1}{2}R g_{MN}$, $R_{MN}$ and $R$ are the Einstein tensor, the Ricci tensor and Ricci scalar, respectively. The ansatz of metric is taken as~\cite{Li:2016gfn}
\begin{equation}
\label{eq3}
\ ds^{2}=\frac{\textit{R}^{2}}{z^{2}}[-f(z)dt^{2}+h(z)(dx_{1}^{2}+dx_{2}^{2})+q(z)dx_{3}^{2}+\frac{dz^{2}}{f(z)}],
\end{equation}
where $f(z = z_{h})=0$ locates at horizon $z=z_h$, and $q(z)$ together with $h(z)$ are regular function of $z$ for $0 \le z \le z_{h}$. The specific equations of motion derived from the action and the perturbative solution of $f(z)$, $h(z)$ and $q(z)$ have been discussed by~\cite{Li:2016gfn}. Using the $r = \frac{\textit{R}^2}{z}$($\textit{R}=1$), we can derive a metric with $r$ as follows:
\begin{equation}
\label{eq4}
\ ds^{2}=r^{2}\left(-f(r)dt^{2}+h(r)(dx_{1}^{2}+dx_{2}^{2})+q(r)dx_{3}^{2}\right)+\frac{1}{r^2 f(r)}dr^{2}.
\end{equation}

As a first order approximation, one can take the leading expansion in Refs.\cite{Li:2016gfn,Zhu:2019ujc} as
\begin{equation}
\label{eq5}
  f(r) = 1- \frac{r_{h}^{4}}{r^{4}} \left( 1-\frac{2 B^{2}}{3 r_{h}^{4}} \log(\frac{r_{h}}{r}) \right ),
 \end{equation}
 \begin{equation}
 \label{eq6}
  q(r) = 1+ \frac{2}{3} B^{2}\log(\frac{1}{r})\frac{1}{r^4},
 \end{equation}
 \begin{equation}
 \label{eq7}
  h(r) = 1- \frac{1}{3} B^{2}\log(\frac{1}{r})\frac{1}{r^4}.
\end{equation}

Noticing that $B$ is related to the physical magnetic field $\mathfrak{B}$ at the boundary by the equation $\mathfrak{B}=\sqrt{3}B$, and $z_h$ is the horizon of the black hole.
If we take $B\leq 0.15 \textrm{GeV}^{2}$, the corresponding physical magnetic field is $\mathfrak{B}\leq 0.26 \textrm{GeV}^{2}$, which conforms to the magnitude of real magnetic field generated by in the RHIC and LHC energies.  The Hawking temperature with magnetic field $B$ is computed as
 \begin{equation}
\label{eq8}
T(z_{h}, B) = \frac{r_{h}}{\pi}-\frac{B^{2}}{6\pi r_{h}^{3}},
\end{equation}

\noindent where $T(z_{h}, B)$  is a function of the position of the horizon and the magnetic field, corresponds to the temperature of the thermal bath in the gauge theory. At the end of the setup, It's necessary to check the validity of the first order pertubative solutions of $f(z)$, $h(z)$ and $q(z)$ in Eqs.(5-7). It was pointed out \cite{Mamo:2015dea} that the perturbative solution can work well only when $B\ll T^{2}$. After inserting the temperature and magnetic field into the IR expansion, Refs.\cite{Li:2016gfn,Zhu:2019ujc} made some comparisons of the leading, next-leading and next-next-leading order of these perturbative solutions, and found that the approximate of leading order Eqs.(5-7) is good enough for $T \geq 0.15~ \textrm{GeV}$ and $B\leq 0.15~ \textrm{GeV}^{2}$. In the paper, the corresponding temperature range and magnetic field range are chosen as $0.15 ~\textrm{GeV} \leq T \leq 0.33~  \textrm{GeV}$ and $0.02 ~\textrm{GeV}^{2} \leq B \leq 0.15 ~\textrm{GeV}^{2}$, which conforms to the range of temperature and magnetic field in the RHIC and LHC erergies.

\section{Dipole moving parallel to the magnetic feild }\label{sec:03}
In the section, we assume that the initial state of $Q\bar{Q}$ is oriented in the direction of magnetic field, and the magnetic field direction is along $x_{3}$ axis. A reference frame is chosen where the plasma is at rest and the $Q\bar{Q}$ dipole is moving with a constant rapidity, and one can boost to a reference frame where the dipole is at rest but the plasma is moving past it. We can utilize this fact to study the effect of the plasma on a $Q\bar{Q}$ pair in the thermal medium. By considering the plasma is at rest, one can boost our frame in one direction with rapidity $\beta$.

When the heavy quark is moving parallel to the magnetic field along the $x_3$ direction with rapidity $\beta$, the coordinates are parametrized by
\begin{equation}
\label{eq9}
\dd{t} = \dd{t}' \cosh \beta - \dd{x_3}' \sinh \beta ,
\end{equation}
\begin{equation}
 \label{eq10}
 \dd{x_3} = -\dd{t}' \sinh \beta + \dd{x_3}' \cosh \beta.
\end{equation}

Substituting (\ref{eq9}) and (\ref{eq10}) into the metric (\ref{eq4}), one can obtain
\begin{gather}
\label{eq11}
\ ds^{2}=\left(-r^{2}f(r) \cosh^2\beta+r^{2}q(r) \sinh^2\beta\right) dt^{2}-2 \sinh\beta \cosh\beta\left(r^2q(r)-r^2f(r)\right)dtdx_3 \nonumber \\
+ \left(-r^2f(r) \sinh^2 \beta+r^2q(r) \cosh^2\beta\right)dx_3^2+r^2h(r)(dx_{1}^2+dx_{2}^2) + \frac{1}{r^2f(r)}dr^2.
\end{gather}

By considering the dipole moving parallel to the wind, one can take
\begin{equation}
\label{eq12}
\ t = \tau, x_1 = x_2 = 0, x_3 = \sigma, r= r(\sigma),
\end{equation}

Then the metric is given as
\begin{gather}
\label{eq13}
\ ds^{2}=(-r^{2}f(r)\cosh^2\beta+r^{2}q(r)\sinh^2\beta)d\tau^{2}  \nonumber \\
         +(-r^2f(r) \sinh^2\beta+r^2q(r)\cosh^2\beta+\frac{\dot{r}^{2}}{r^2f(r)})d\sigma^2.
\end{gather}

Holographically, in the supergravity limit which corresponds to a strongly coupled plasma, one can evaluate the expectation value of Wilson loop $W(C)$ by the prescription
\begin{equation}
\label{eq14}
\    \langle W(C) \rangle\sim e^{-iS_{NG}},
\end{equation}

\noindent where $S_{NG}$ is the classical Nambu-Goto action of a string in the bulk, which can be given as:
\begin{equation}
\label{eq15}
\    S_{NG} = - \frac{1}{2\pi\alpha'}\int \dd\sigma d\tau \sqrt{- \det g_{ab}},
\end{equation}

\noindent where the induced metric of the worldsheet $g_{ab}$  is given by
\begin{equation}
\label{eq16}
\   g_{ab} = g_{MN} \partial_a X^M \partial_b X^N, \quad a,\,b=0,\,1,
\end{equation}

\noindent evaluated at an extremum of the action, $\delta S_{NG} = 0$. The string worldsheet coordinates are given in static gauge, $X^{\mu}=(t,x_{3},0,0,r(x_{3},t))$,
$\tau = t$ and $\sigma=x_{3}$ for the rectangular Wilson loop.

If the quark and anti-quark are located at $x_3 = L/2$ and $x_3 = -L/2$, respectively, the rectangular Wilson loop of spatial length $L$ and time extension $\mathcal{T}$ are studied. Plugging back $S_{NG}$ in (\ref{eq14}) one can evaluate the real part of $V_{Q\bar{Q}}$. By
considering thermal fluctuations of the string, one will be able to evaluate the imaginary part of $V_{Q\bar{Q}}$.
With the metric of (\ref{eq13}), the Nambu-Goto action (\ref{eq15}) can be calculated as
\begin{equation}
\label{eq17}
\    S_{NG} =  \frac{\mathcal{T}}{2\pi\alpha'}\int_{-L/2}^{L/2} \dd x_{3} \sqrt{a(r)+b(r)\dot{r}^{2}}.
\end{equation}

The Lagrangian density is taken as:
\begin{equation}
\label{eq18}
\mathcal{L} = \sqrt{a(r)+b(r)\dot{r}^{2}},
\end{equation}

\noindent with
\begin{equation}
\label{eq19}
\ a(r) = r^4{f(r)q(r)(\cosh^4\beta+ \sinh^4\beta)-\sinh^2\beta \cosh^2\beta(f^{2}(r)+q^{2}(r))},
\end{equation}

\noindent and
\begin{equation}
\label{eq20}
\ b(r) = \cosh^2\beta-\frac{q(r)}{f(r)} \sinh^2(\beta).
\end{equation}

Note that the Lagrangian density does not depend on $x_{3}$ explicitly, then a conserved quantity can be given as:
\begin{equation}
\label{eq21}
\mathcal{L} - \frac{\partial \mathcal{L}}{\partial \dot{r}} \dot{r} = constant
\end{equation}

The boundary condition at $x_{3} = 0 $, when $r = r_c$, $\dot{r}_{c} ̇= dr/dx_{3}\mid_{r=r_{c}} = 0$.
From above, we can derive
\begin{equation}
\label{eq22}
\  \dot{r} = \sqrt{ \frac{a(r)^2-a(r_c)a(r)}{a(r_c)b(r)}}.
\end{equation}

The distance of the heavy $Q\bar{Q}$ pair can be calculated as
\begin{equation}
\label{eq23}
\  L = 2 \int^{\infty}_{r_c} dr \sqrt{ \frac{a(r_c)b(r)}{a(r)^2-a(r_c)a(r)}}.
\end{equation}

The real part of the heavy quark potential can be derived as:
\begin{equation}
\label{eq24}
\ \textrm{Re}V_{Q\bar{Q}} = \frac{1}{\pi \alpha'} \int^{\infty}_{r_c} dr\left(\sqrt{ \frac{a(r)b(r)}{a(r) - a(r_c)}}-\sqrt{b_{0}(r)}\right) - \frac{1}{\pi \alpha'} \int^{r_c}_{r_h} dr \sqrt{b_{0}(r)},
\end{equation}

\noindent where $b_{0}(r)=b(r\rightarrow\infty)$.

\begin{figure}[H]
    \centering
    \includegraphics[width=0.8\textwidth,natwidth=610,natheight=642]{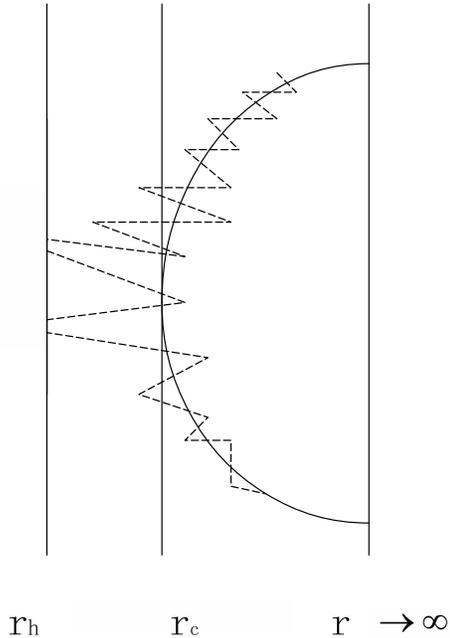}
    \caption{\label{fig1}  An illustration of the effect of thermal fluctuations around the classical string configuration (solid line). The dashed line is for thermal fluctuations.}
\end{figure}

The real part of the heavy quark potential as (\ref{eq24})  can be computed by using the classical solution to the Nambu-Goto action (\ref{eq17}). To explore $\textrm{Im}V_{Q\bar{Q}}$ we have to consider the effect of thermal worldsheet fluctuations of the classical configuration $r = r_{c}(x_{3})$. Although such fluctuations should be small, but they may turn the integrand of (\ref{eq17}) negative near $x_{3} = 0$ and create an imaginary part for the effective string action. The corresponding physical picture is that some part of the string may reach the horizon through thermal fluctuations shown in Fig~\ref{fig1}. If the bottom of the classical string solution is close enough to the horizon, thermal worldsheet fluctuations of very long wavelength may be able to reach the black brane horizon.
It is well known that an imaginary potential $\textrm{Im}V_{Q\bar{Q}}$ is an imaginary part of the potential, which can be used to define a thermal decay width. For weak coupling the thermal width is associated with the imaginary part of the gluon self-energy induced by Landau damping and the $Q\bar{Q}$ color singlet-to-octet thermal break-up.
In this approach, the thermal width of heavy quarkonium states comes from the effect of  the thermal fluctuation due to the interactions between the heavy quarks and the strongly coupled medium. By considering the thermal worldsheet fluctuations about classical configuration, one can extract imaginary potential and thermal width, the detailed analysis can be found in~\cite{Finazzo:2013aoa,Fadafan:2013bva}. The imaginary potential is given as:
\begin{gather}
\label{eq25}
\ \textrm{Im}V_{Q\bar{Q}} = -\frac{1}{2\sqrt{2} \alpha'} \left(\frac{\sqrt{a'(r_c)}}{2a''(r_c)} - \frac{a(r_c)}{a'(r_c)}\right)\sqrt{b_{0}(r)},
\end{gather}

\noindent where $a'(r_c)$ and $a''(r_c)$ are the values of the first and second derivative of $a(r)$ to $r$ at $r_c$, respectively.

As follow, we will use a first-order non-relativistic expansion~\cite{Finazzo:2013aoa}~to estimate the thermal width $\Gamma_{Q\bar{Q}}$ of the heavy quarkonia
\begin{equation}
\label{eq26}
\ \Gamma_{Q\bar{Q}} = -\langle\psi\mid \textrm{Im}V_{Q\bar{Q}}(L, T)\mid\psi\rangle,
\end{equation}
where
\begin{equation}
\label{eq27}
\ \langle\vec{r}|\psi\rangle = \frac{1}{\sqrt{\pi}a_{0}^{3/2}}e^{-r/a_{0}}
\end{equation}
is the ground-state wave function of a particle in a Coulomb-like potential $V(L) = -K/L$, where the Bohr radius is defined as $a_{0} = 2/(m_Q K)$,  $m_{Q}$ is the mass of the heavy quark $Q$. For the $\Upsilon(1S)$ state, the thermal width is given as
\begin{equation}
\label{eq28}
\ \frac{\Gamma_{Q\bar{Q}}}{T} = -\frac{4}{Ta_0} \int^{\infty}_{0} d\omega e^{-\frac{2\omega}{Ta_0}}\omega^2\frac{\textrm{Im}V_{Q\bar{Q}}}{T}(\omega),
\end{equation}
where $a_0 \sim 0.6$ GeV$^{-1}$ and $m_{Q} = 4.6 \textrm{GeV}$ for the calculation of $\Upsilon(1S)$ thermal width. Note that the imaginary potential is defined in the region $(L_{min}, L_{max})$ instead of taking the integral from zero to infinity, which means the imaginary potential starts at a $L_{min}$ which can be computed by solving $\textrm{Im}V_{Q\bar{Q}} = 0$ and ends at a $L_{max}$. The solution is called as a conservative approach~\cite{Fadafan:2015kma}.

\section{Dipole moving perpendicular to the magnetic feild }\label{sec:04}
In this section, we assume that the initial state of $Q\bar{Q}$ is oriented transverse to the direction of magnetic field, and the magnetic field direction is along $x_{3}$ axis. By considering the system moves along the $x_{1}$ direction with rapidity $\beta$, one can take
\begin{equation}
\label{eq29}
\ t = \tau, x_2 = x_3 = 0, x_1 = \sigma, r= r(\sigma).
\end{equation}

Then the metric of dipole moving perpendicular to the magnetic field becomes
\begin{gather}
\label{eq30}
\ ds^{2}=\left(-r^{2}f(r)\cosh^2\beta+r^{2}q(r)\sinh^2\beta\right)dt^{2}+\left(r^2h(r)+\frac{\Dot{r}^2}{r^2f(r)}\right)dx_{1}^{2} .
\end{gather}

The Lagrangian density is
\begin{equation}
\label{eq31}
\   \mathcal{L}_{P} = \sqrt{a_{P}(r)+b_{P}(r)\dot{r}^{2}},
\end{equation}
where
\begin{equation}
\label{eq32}
\ a_{P}(r) = r^{4}f(r)h(r)\cosh^2\beta-r^{4}q(r)h(r)\sinh^2\beta,
\end{equation}
and
\begin{equation}
\label{eq33}
\ b_{P}(r) = \cosh^2\beta-\frac{q(r)}{f(r)}\sinh^2(\beta).
\end{equation}

The real part of the heavy quark potential can be calculated as
\begin{gather}
\label{eq34}
\ \textrm{Re} V_{Q\bar{Q}} = \frac{1}{\pi \alpha'} \int^{\infty}_{r_c} dr\left(\sqrt{ \frac{a_{P}(r)b_{P}(r)}{a_{P}(r) - a_{P}(r_c)}}-\sqrt{b_{P0}(r)}\right) - \frac{1}{\pi \alpha'} \int^{r_c}_{r_h} dr \sqrt{b_{P0}(r)}.
\end{gather}
where $b_{P0}(r)=b_{P}(r\rightarrow\infty)$.

Similarly, one can calculate the imaginary part of the heavy quark potential when dipole moving perpendicular to the magnetic field
\begin{equation}
\label{eq35}
\ \textrm{Im} V_{Q\bar{Q}} = -\frac{1}{2\sqrt{2} \alpha'} \left(\frac{\sqrt{a_{P}'(r_c)}}{2a_{P}''(r_c)} - \frac{a_{P}(r_c)}{a_{P}'(r_{c})}\right)\sqrt{b_{P0}(r)}.
\end{equation}

\section{Comparison of dipole moving  parallel and perpendicular to the magnetic field}\label{sec:05}
\begin{figure}
    \centering
    \includegraphics[width=10cm]{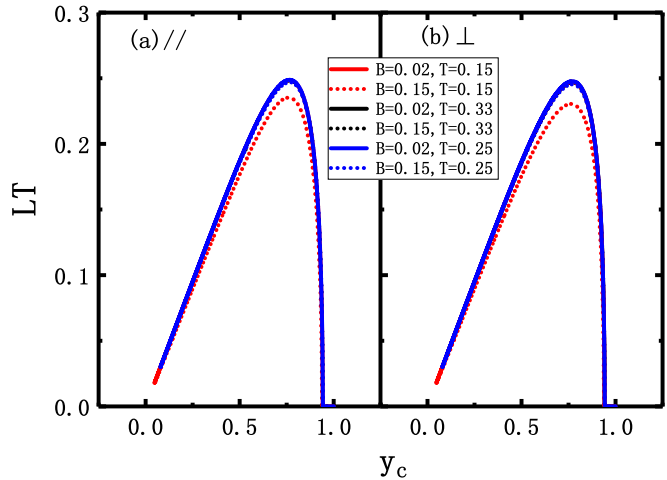}
    \caption{\label{fig2} LT versus $y_c$ when dipole moving parallel and perpendicular to the magnetic fields for different temperatures and magnetic fields at fixed moving rapidity($\beta = 0.5$). Fig.2(a)and Fig.2(b) are for the case of dipole moving parallel and perpendicular to the magnetic field, respectively. Solid line is for $B = 0.02 \textrm{GeV}^2$ and dashed line is for $B = 0.15 \textrm{GeV}^2$.}
\end{figure}

In order to study the effects of the magnetic field on the imaginary potential and $\Upsilon(1S)$ thermal distributions in finite temperature magnetized background, we make a comparison of dipole moving parallel and perpendicular to the magnetic fields cases. As we know, the $Q\bar{Q}$ pair is not generated in static QGP medium, but in moving QGP medium. Therefore, we should consider the influence of the moving QGP medium on the $Q\bar{Q}$ pair. In order to study quarkonia thermal features in RHIC and LHC energies, we choose some special magnetic fields $B = 0.02~ \textrm{GeV}^2$, $B = 0.15~ \textrm{GeV}^2$ and with temperatures as $T = 0.15, 0.25$ and $0.33~ \textrm{GeV}$ in RHIC and LHC energies. Fig.~\ref{fig2} publishes the behavior of $LT$ as a function of $y_{c}=r_{h}/r_{c}$ for different magnetic fields $B$ with different temperatures. It is found that the magnetic field has a significant effect on the relationship between $LT$ and $y_{c}$ at the low deconfined temperature ($T_{c}=0.15 \textrm{GeV}$). But as the increase of temperature $T > T_{c}$, the influence of the magnetic field on the relationship between $LT$ and $y_{c}$
becomes less and less significant whether dipole moving parallel or perpendicular to the magnetic field.  There is a maximum value of $LT$ locating at $y_{c,max}$, when $y<y_{c,max}$, $LT$ is an increasing function of $y_c$, but when $y>y_{c,max}$, $LT$ is a decreasing function of $y_c$.

\begin{figure}
    \centering
    \includegraphics[width=10cm]{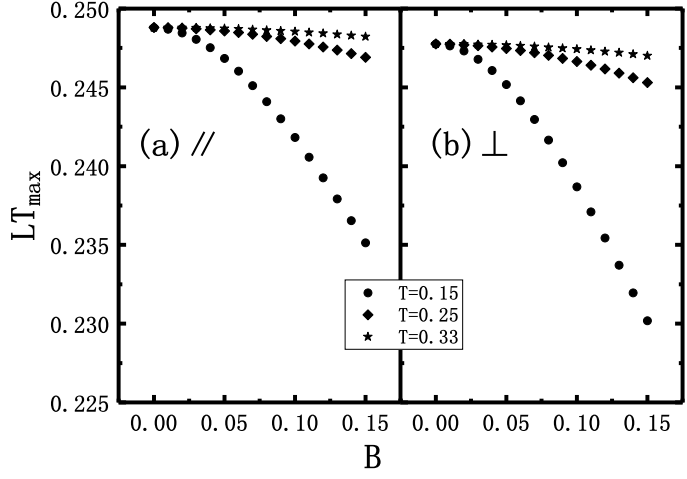}
    \caption{\label{fig3} The dependencies of maximum value of $LT$ ($LT_{\textrm{max}}$) on magnetic field $B$ with different temperatures for dipole moving parallel(a) and perpendicular(b) to the magnetic field direction, respectively}
\end{figure}

The maximum values of $LT$ ($LT_{\textrm{max}}$), which defines a dissociation length for $Q\bar{Q}$\cite{Liu:2006nn,Liu:2006he}, as a function of magnetic field $B$ has been studied in Fig.~\ref{fig3}. Fig.~\ref{fig3} illustrates that increasing $B$ reduces $LT_{\textrm{max}}$ for the moving $Q\bar{Q}$.
It is found that the magnetic field has a significant effect on the dissociation length for $Q\bar{Q}$ at the low deconfined temperature $T_{c}$, however, with the increase of temperature, the effect of magnetic field on dissociation length $LT_{\textrm{max}}$ become less and less significant. When the temperature reaches $0.33 \textrm{GeV}$, $LT_{\textrm{max}}$ remains unchanged with the increase of magnetic field. In this case, the dominant configuration for $S_{NG}$  should be two straight strings running from the boundary to the horizon. Moreover, the dissociation properties of heavy quarkonia should be sensitive to the imaginary part of the potential.

\begin{figure}
    \centering
    \includegraphics[width=10cm]{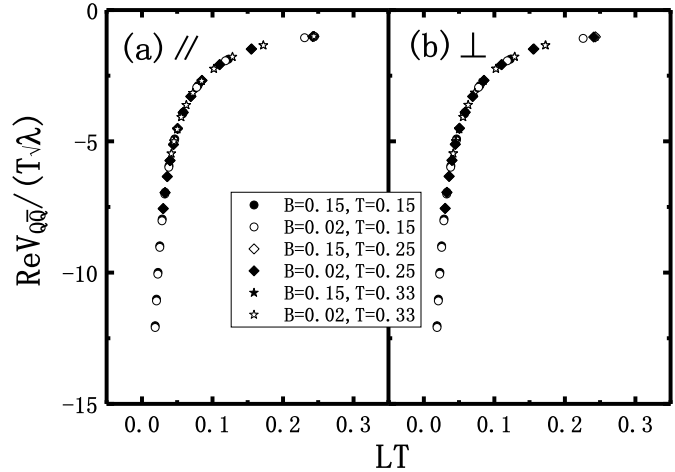}
    \caption{\label{fig4} The real potential of quark-antiquark pair versus $LT$ with different magnetic fields and temperatures for dipole moving parallel(a), and perpendicular (b) to magnetic field, respectively}
\end{figure}

Fig.~\ref{fig4}(a, b) include comparisons of the real potential of $Q\bar{Q}$ pair versus $LT$  for a dipole moving parallel and perpendicular to the magnetic field, respectively.
Generally speaking, the different of the effects of dipole moving parallel, and perpendicular to magnetic fields on the relationship between real potential and $LT$ is not very obvious. From Fig.4(a, b), we find out that no matter how the magnetic field or temperature changes, the real potential remains unchanged in the RHIC and LHC energy regions.

\begin{figure}
    \centering
    \includegraphics[width=10cm]{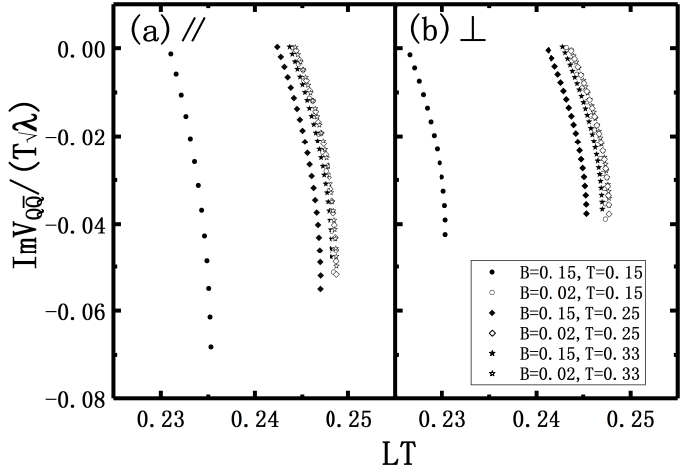}
    \caption{\label{fig5} The imaginary potential $\textrm{Im}V_{Q \bar{Q}}$ of quark-antiquark pair versus $LT$ with different magnetic field and temperature for dipole moving parallel (a), and perpendicular (b) to the magnetic field,respectively. Solid line is $B = 0.02$ GeV $^2$, and dashed line is $B = 0.15$GeV$^2$ .}
\end{figure}

As shown in Fig.~\ref{fig5}(a, b), The imaginary potential starts at a $L_{min}$ which can be computed by solving
$\textrm{Im}V_{Q\bar{Q}} = 0$ and ends at a $L_{max}$.  We also find that at the low deconfined temperature ($T =T_{c}= 0.15~ \textrm{GeV}$) increasing the magnetic field leads to an increase of the absolute value of the imaginary potential, and makes the imaginary potential $\textrm{Im}V_{Q \bar{Q}}$ occurring at a smaller inter-quark distance. When the temperature increases to $T = 0.33 ~\textrm{GeV}$, the effect of the magnetic field on $\textrm{Im}V_{Q \bar{Q}}$ becomes less and less important. Comparing Fig. 5 (a) with Fig. 5 (b), we find that the absolute value of the imaginary potential in the case of dipole moving parallel to the magnetic field is apparently bigger than the case of dipole moving perpendicular to the magnetic field.

Fig.6 shows the thermal width of the $\Upsilon(1S)$ with non-zero magnetic fields at different finite temperatures for dipole moving parallel and perpendicular to the magnetic field case. We find that with increasing magnetic field leads to an significant increase of thermal width at the low deconfined temperature ($T = 0.15 \textrm{GeV}$). But with the increase of temperature, the effect of the magnetic field on thermal width becomes less and less significant Comparing Fig. 6 (a) with Fig. 6 (b), we find that the thermal width in the case of dipole moving parallel to the magnetic field is apparently bigger than the case of dipole moving perpendicular to the magnetic field.
It is well known that the imaginary potential can be utilized to assess the thermal width of heavy quarkonium. It is argued that a large thermal width corresponds to a large dissociation length. Thus, the magnetic field has the effect of increasing the dissociation length or increasing the thermal width  at the low deconfined temperature ($T = T_{c} = 0.15~\textrm{GeV}$). However, with the increase of temperature, the effect of magnetic field on thermal width and the dissociation length will gradually weaken.
\begin{figure}[H]
    \centering
    \includegraphics[width=11cm]{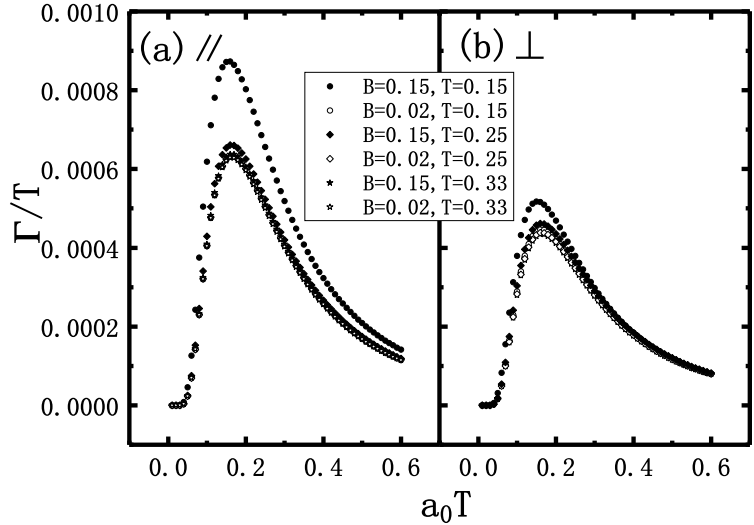}
    \caption{\label{fig6}  The thermal width $\Gamma_{Q \bar{Q}}$ of $\Upsilon(1S)$ state divided by the temperature $T$ versus $Ta_{0}$ for different magnetic fields and temperatures for dipole moving  parallel(a) and perpendicular(b) to the magnetic field, respectively. The fixed rapidity is given as $\beta = $ 0.5.}
\end{figure}

Fig.7 shows the thermal width of the $\Upsilon(1S)$ with different rapidity and temperatures for dipole moving parallel and perpendicular to the magnetic field case, respectively. It is found that the thermal width decreases as the increasing rapidity at a fixed temperature, which is in agreement with that computed by Refs.\cite{Escobedo:2011ie, Ali-Akbari:2014vpa}.  The thermal width of $\Upsilon(1S)$ hardly changes with temperature at a fixed rapidity and magnetic field in the case of dipole moving perpendicular to the magnetic field as shown in Fig.7(b). But in the case of dipole moving parallel to the magnetic field as shown in Fig.7(a), the thermal width at low deconfined temperature ($T = 0.15~ \textrm{GeV}$) is obviously larger than that at high temperature.
\begin{figure}[H]
    \centering
    \includegraphics[width=11cm]{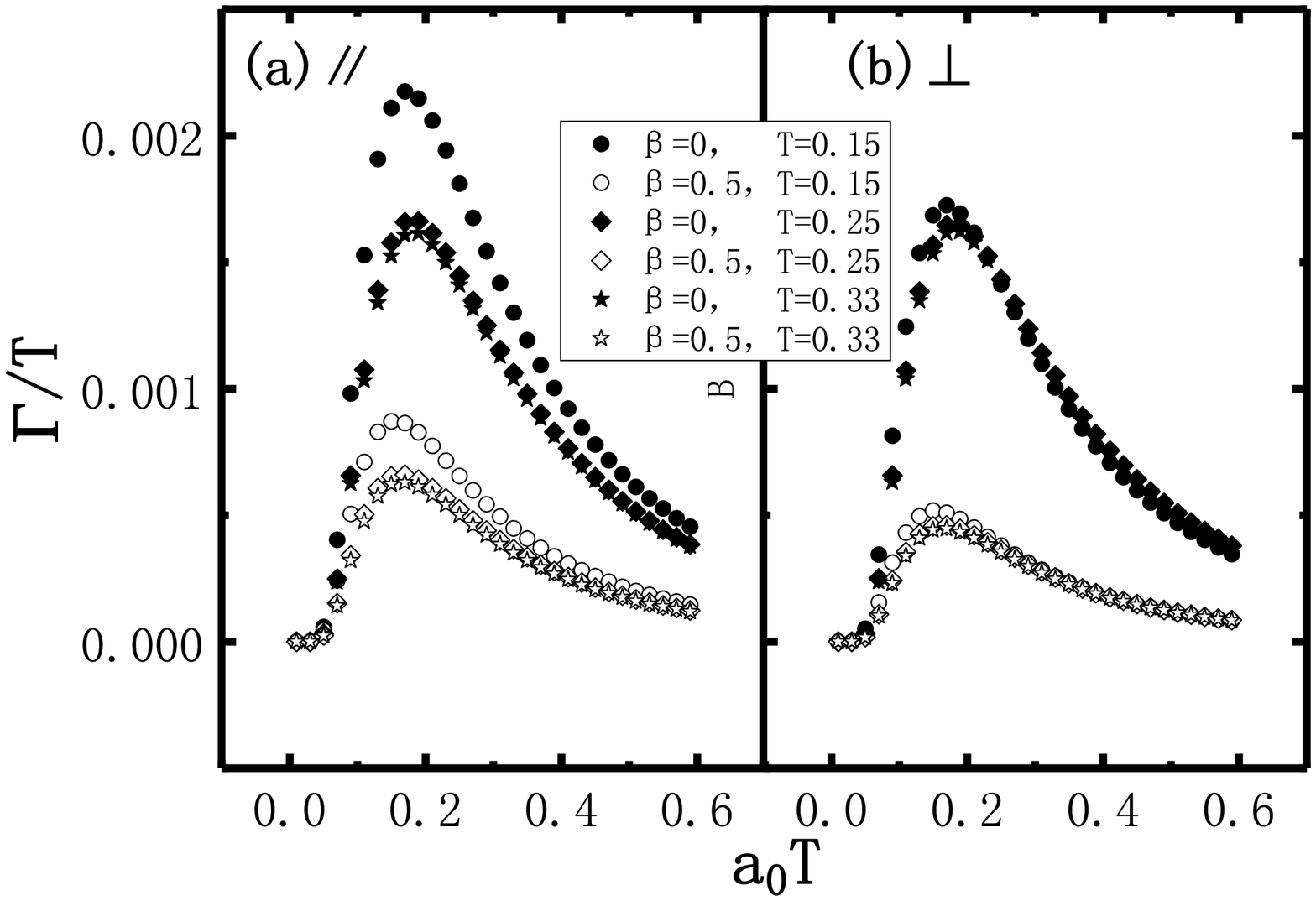}
    \caption{\label{fig7} The thermal width $\Gamma_{Q \bar{Q}}$ of $\Upsilon(1S)$ state divided by the temperature $T$ versus $Ta_{0}$ for different rapidity and temperature for dipole moving parallel(a) and perpendicular(b) to the magnetic field, respectively. The fixed magnetic field is given as $B = 0.15 \textrm{GeV}^{2}$. }
\end{figure}

\section{SUMMARY AND CONCLUSION}\label{sec:06}
The heavy-quark potential, whether it is a real potential or a imaginary potential, both are very important quantity in gauge theories at finite temperature. It also has great
relevance in connection with experimental programs in heavy ion collisions in the RHIC and LHC energies. The melting of heavy quarkonia in a medium is considered to be one of the main experimental signatures for QGP formation. Current analyses of available researches
indicate that the matter formed in such collisions is strongly coupled. Thus, the study of the heavy quark potential requires strong-coupling techniques, such
as the AdS/CFT correspondence. There has been a lot of interests on the heavy quarkonium suppression which has been observed in the
RHIC\cite{Adams:2005dq} and LHC~\cite{Chatrchyan:2012lxa,Abelev:2012rv}. The suppression is a signal of deconfinement and it was suggested that the bound states
dissociate in the hot thermal bath. It was proposed that the imaginary part of the potential $\textrm{Im}⁡V_{Q\bar{Q}}$ may be an important reason responsible for this suppression rather than color screening.

By simulating the finite temperature  and magnetic field in the RHIC and LHC energy regions of relativistic heavy ion collisions, we restrict ourselves to the range of temperature and magnetic field corresponding to RHIC and LHC energy regions to study the potential and thermal width for dipole moving parallel and pedicular to the magnetic field.
It is found that the magnetic field has less influence on the real potential, but has greater influence on the imaginary potential. Extracting from the effect of thermal worldsheet fluctuations about the classical configuration, we investigate the
thermal width of $\Upsilon(1s)$ in the finite temperature magnetized background. The thermal width of $\Upsilon(1s)$ increases with the increasing magnetic field at the low deconfined temperature ($T_{c}=0.15 \textrm{GeV}$), but with the increase of temperature ($T>T_{c}$), the thermal width hardly changes with the increase of magnetic field, which means the effect of high temperature completely exceeds that of magnetic field. The thermal width decreases with the increasing rapidity at the finite temperature magnetized background.

It is also found that the effects of a magnetic field on the thermal width when dipole moving parallel to the magnetic field direction are larger than dipole moving perpendicular to the magnetic field direction, which implies that the magnetic field tends to enhance thermal fluctuation when dipole moving parallel to the magnetic field. The thermal width of $\Upsilon(1S)$ hardly changes with temperature at fixed rapidity and magnetic field in the case of dipole moving perpendicular to the magnetic field. But in the case of dipole moving parallel to the magnetic field, the thermal width at low temperature ($T = 0.15~ \textrm{GeV}$) is obviously larger than that at high temperature.

\section*{Acknowledgments}
This work was supported by National Natural Science Foundation of China (Grants No. 11875178, No. 11475068, and No. 11747115) and the CCNU-QLPL Innovation Fund (Grant No. QLPL2016P01).

\end{document}